\documentstyle[12pt]{article}

\begin{document}

\baselineskip=18pt plus 0.2pt minus 0.1pt
\parskip = 6pt plus 2pt minus 1pt

\catcode`\@=11

\newif\iffigureexists
\newif\ifepsfloaded
\openin 1 epsf.sty
\ifeof 1 \epsfloadedfalse \else \epsfloadedtrue \fi
\closein 1
\ifepsfloaded
    \input epsf.sty
\else
    \immediate\write20{>Warning:
         No epsf.sty --- cannot embed Figures!!}
\fi
\def\checkex#1 {\relax
    \ifepsfloaded \openin 1 #1
        \ifeof 1 \figureexistsfalse
        \else \figureexiststrue
        \fi \closein 1
    \else \figureexistsfalse
    \fi }

\def\epsfhako#1#2#3#4#5#6{
\checkex{#1}
\iffigureexists
    \begin{figure}[#2]
    \epsfxsize=#3
    \centerline{\epsffile{#1}}
    {#6}
    \caption{#4}
    \label{#5}
    \end{figure}
\else
    \begin{figure}[#2]
    \caption{#4}
    \label{#5}
    \end{figure}
    \immediate\write20{>Warning:
         Cannot embed a Figure (#1)!!}
\fi
}

\ifepsfloaded
\checkex{betac.eps}
    \iffigureexists \else
    \immediate\write20{>EPS files for Figs. 1 to 5 are packed
     in a uuencoded compressed tar file}
    \immediate\write20{>appended to this LaTeX file.} 
    \immediate\write20{>You should unpack them and LaTeX again!!}
    \fi
\fi

\begin{titlepage}
\title{
\hfill
\parbox{4cm}{\normalsize OPCT-94-1\\hep-lat/9407023}\\
\vspace{1cm}
         Low-Temperature Series for the Correlation Length  
         in $d=3$ Ising Model 
       }
\author{
        H. ARISUE \\ 
        {\normalsize\em Osaka Prefectural College of Technology}\\ 
        {\normalsize\em Saiwai-cho, Neyagawa, Osaka 572, Japan}
              \and
        K. TABATA \\
        {\normalsize\em Osaka Institute of Technology}\\ 
        {\normalsize\em Junior College}\\ 
        {\normalsize\em Ohmiya, Asahi-ku, Osaka 535, Japan}
        }
\date{\normalsize July, 1994}
\maketitle
\thispagestyle{empty}

\begin{abstract}
\normalsize
  We extend low-temperature series 
for the second moment of the correlation function 
in $d=3$ simple-cubic Ising model 
from $u^{15}$ to $u^{26}$ using finite-lattice method, 
and combining with the series for the susceptibility 
we obtain the low-temperature series 
for the second-moment correlation length 
to $u^{23}$. 
  An analysis of the obtained series 
by inhomogeneous differential approximants 
gives critical exponents  
$ 2\nu^{\prime} + \gamma^{\prime} \approx 2.55 $ 
and 
$ 2\nu^{\prime} \approx 1.27 $. 
\end{abstract}
\end{titlepage}

\newpage
\section{Introdunction}
\hspace*{\parindent}

 The low-temperature series of $d=3$ Ising model 
or equivalently strong coupling series of $d=3$ $Z_2$ 
lattice gauge theory had long been shorter 
than high-temperature series.  
Recently they have been extended 
to higher orders 
          \cite{Arisuethree_five,Bhanot,Guttmann,Arisuefour}
using finite-lattice method 
           \cite{Neef,Arisueone_two_six,Creutz}.   
 In this method free energy density in the infinite volume limit, 
for instance, is given by a linear combination of 
the free energy on appropriate finite-size lattices. 
 The algorithm to give the coefficients of 
the linear combination is so simple. 
 In the standard graphical method, 
it is rather difficult 
to list up all the diagrams completely 
that contribute to the relevant order of the series. 
  The finite-lattice method avoids this problem 
involved in the standard graphical method 
and enables us to obtain longer series. 
  The method was applied in $d=3$ Ising model 
to get the low-temperature series of 
the true inverse correlation length 
(which is equivalent to the mass gap in lattice gauge theory) 
          \cite{Arisuethree_five},
free energy 
          \cite{Bhanot},
magnetization and zero-field susceptibility
          \cite{Guttmann},
and surface tension 
(which is equivalent to the string tension in lattice gauge theory) 
          \cite{Arisuefour}.

 In this paper we apply the method 
to calculate the low-temperature expansion series for 
the second moment of the correlation function $\mu_2$
in $d=3$ simple-cubic Ising model to order $u^{26}$, 
extending the previous result of order $u^{15}$ by Tarko and Fisher
          \cite{Tarko}. 
 It gives the low-temperature series 
for the second-moment correlation length squared $\Lambda_2={\xi_1}^2$
          \cite{Tarko} 
to order $u^{23}$, 
when combined with the low-temperature series of the susceptibility.
 This is longer by five terms 
than the low-temperature series for true correlation length 
squared $\Lambda_2^{\prime}$ 
that was derived from the true inverse correlation-length 
given in Ref.~\cite{Arisuethree_five}.

 In the next section we present the algorithm 
to obtain low-temperature expansion series for $\mu_2$ 
using finite-lattice method. 
 In section 3 the expansion series for $\mu_2$ and $\Lambda_2$ 
is given. The low-temperature series for true correlation length 
squared $\Lambda_2^{\prime}$ is also listed for comparison. 
 The result of series analysis 
by inhomogeneous differential approximants 
is described in section 4.

\section{ Algorithm of low-temperature expansion }
\hspace*{\parindent}

 The second-moment correlation length squared is defined by
\begin{equation}
      \Lambda_2 = \frac{\mu_2}{2 d \mu_0},
\end{equation}
where 
$\mu_2$ is the second moment of the correlation function 
\begin{equation}
   \mu_2 = 
       \lim_{V \rightarrow \infty}     \frac{1}{V} 
  \sum_{i,j} ( \mbox{\boldmath$r$}_i - \mbox{\boldmath$r$}_j )^2 
       \langle s_i s_j \rangle_c ,
\end{equation}
with
$V$ the lattice volume 
and $\mbox{\boldmath$r$}_i =( x_i , y_i, z_i )$ 
the coordinate of the lattice site $i$,  
and $\mu_0$ is the zero-th moment of the correlation function 
or the susceptibility 
and $d$ is the dimensionality of the lattice. 
 Here in this paper we take the lattice spacing $a=1$.

 The algorithm to calculate the low-temperature expansion 
of the second moment $\mu_2$ is the following. 
 We consider the partition function 
\begin{equation}
   Z(\beta,h,\eta,\gamma_1,\gamma_2,\gamma_3)
       = \sum_{ \{ s_i \} } \exp{( - {\cal H} )},    \label{eqn:Z}
\end{equation}
with Hamiltonian 
\begin{equation}
    {\cal H } = \beta \sum_{\langle ij \rangle} s_i s_j 
            + \sum_i ( h + \gamma_1 x_i + \gamma_2 y_i + \gamma_3 z_i 
                          + \eta \mbox{\boldmath$r$}_i^2 )  s_i, 
\end{equation}
for a three-dimensional rectangular lattice $\Lambda_0$ 
with a volume $ V = L_x \times L_y \times L_z $. 
 We take the fixed boundary condition 
that all the spins outside $\Lambda_0$ are aligned, 
for instance, to be $\{ s_i=+1 \}$. 
 Then the second moment $\mu_2$ is given by 
\begin{equation}
   \mu_2 =
       \lim_{V \rightarrow \infty}     \frac{2}{V} 
    \left( \frac{\partial^2}{\partial h \partial \eta } 
        -  \frac{\partial^2}{\partial \gamma_1^2 } 
        -  \frac{\partial^2}{\partial \gamma_2^2 } 
        -  \frac{\partial^2}{\partial \gamma_3^2 }  \right) 
             \ln{ Z(\beta,h,\eta,\gamma_1,\gamma_2,\gamma_3) }
                |_{h=\eta=\gamma_1=\gamma_2=\gamma_3=0} .
\end{equation}
 Let us consider the set $\{\Lambda\}$ 
of all three-dimensional rectangular sub-lattices of $\Lambda_0$ 
$( \Lambda \subseteq  \Lambda_0 )$ 
with the volume $l_x \times l_y \times l_z$
and define $H$ of $\Lambda$ as
\begin{equation}
   H(\Lambda) 
     = 
   2 \left( \frac{\partial^2}{\partial h \partial \eta } 
        -  \frac{\partial^2}{\partial \gamma_1^2 } 
        -  \frac{\partial^2}{\partial \gamma_2^2 } 
        -  \frac{\partial^2}{\partial \gamma_3^2 }  \right) 
              \ln{ Z(\Lambda)|_{h=\eta=\gamma_1=\gamma_2=\gamma_3=0} },
\end{equation}
where $Z(\Lambda)$ is the patition function for $\Lambda$ 
with the fixed boundary condition 
that all the spins outside $\Lambda$ is aligned, 
and define $W$ of $\Lambda$ recursively as
\begin{equation}
      W(\Lambda) 
       = H(\Lambda)
         - \sum_{ \Lambda^{\prime} \subset \Lambda } 
          W( \Lambda^{\prime} ) .
\end{equation}
 Note that $H(\Lambda)$ and $W(\Lambda)$ depend 
not on the position but on the size $l_x,l_y,l_z$ of $\Lambda$. 
 We know 
\begin{equation}
      H(\Lambda_0)=
          \sum_{ \Lambda \subseteq \Lambda_0 } 
                  W( \Lambda ).
\end{equation}
 Taking the infinite volume limit we obtain 
\begin{eqnarray}
   \mu_2 
     &=&
       \lim_{ V \rightarrow \infty} 
          \frac{1}{V} 
                 H(\Lambda_0)
                                                        \nonumber \\
     &=&
       \lim_{ L_x,L_y,L_z \rightarrow \infty} 
        \frac{1}{L_x L_y L_z} 
           \sum_{l_x,l_y,l_z} (L_x-l_x+1)(L_y-l_y+1)  \nonumber \\
    & &  \qquad \qquad \qquad \qquad \qquad \qquad
         \times (L_z-l_z+1)  W( l_x,l_y,l_z )         \nonumber \\
    &=&
       \sum_{l_x,l_y,l_z} 
                 W( l_x,l_y,l_z ) .                     \label{eqn:B}
\end{eqnarray}

 We can prove 
       \cite{Arisueone_two_six} 
that the Taylor expansion of $W(\Lambda)$ 
with respect to $u=\exp{(-4\beta)}$ 
includes the contribution from all the clusters of polymers 
in standard cluster expansion 
           \cite{Muenster}
that can be embedded into the rectangular lattice $\Lambda$ 
but cannot be embedded into any of its rectangular sub-lattice 
$\Lambda^{\prime} ( \subset \Lambda )$. 
 The series expansion of $W( \Lambda )$ 
starts from the order of $u^n$ with 
$  n = 2 (l_x + l_y + l_z ) - 3 $. 
 So we should take all the finite-size rectangular lattices 
that satisfy 
$ 2 (l_x + l_y + l_z ) -3 \le N $
for the summation in eq.~(\ref{eqn:B}) 
to obtain the expansion series to order $u^N$.

 In practice for the calculation of 
$\frac{\partial^2}{\partial h \partial \eta } 
   \ln{ Z(l_x,l_y,l_z) }|_{h=\eta=\gamma_1=\gamma_2=\gamma_3=0}$,
for instance, 
we have only to calculate the partition function 
$Z(\beta,h,\eta;\gamma_1=\gamma_2=\gamma_3=0)$ 
to order $h \eta u^N$.

\section{Expansion series}
\hspace*{\parindent}

 The low-temperature series of $\mu_2$ have been obtained 
to order $u^{26}$ 
using rectangular lattices with cross-section up to $4 \times 5$. 
 Bhanot's algorithm of calculating the exact partition function 
for finite-size lattices 
         \cite{Bhanottwo}
can be applied to the partition function (\ref{eqn:Z}), 
in which the necessary memory and CPU time are proportional 
to $ N \times 2^{l_x \times l_y} $
and $ N \times 2^{l_x \times l_y} \times l_x \times l_y \times l_z $,
respectively. 
 The calculation was performed on FACOM-VP2600 
at Kyoto University Data Processing Center 
and HITAC-S820 at KEK, 
both of which have about 0.5 Gbyte of main memory 
and $1 - 2 $ Gbyte of extended storage. 
Total CPU time necessary was about 3 hours.

 The low-temperature series obtained 
is listed in table~\ref{tab:coefficients},  
where the coefficients $\{ m_n \}$ are defined by 
\begin{equation}
   \mu_2 =  \sum_n m_n u^n , 
\end{equation}
\begin{table}[htb]
\caption{
The low-temperature expansion coefficients $\{ m_n \}$
for the second moment of the correlation function $\mu_2$, 
$\{ \lambda_n \}$ 
for the second-moment correlation length squared $\Lambda_2$, 
and $\{ \lambda_n^{\prime} \}$ 
for the true correlation length squared $\Lambda_2^{\prime}$
in $d=3$ simple-cubic Ising model.
         }
\label{tab:coefficients}
\begin{center}
\begin{tabular}{rrrrr}                       \hline
 $n$ &           $m_n$   &  $\lambda_n$ & $\lambda_n^{\prime}$ \\ \hline
  0  & $              0$   & $           0$ & $        0$  \\    
  1  & $              0$   & $           0$ & $        0$  \\    
  2  & $              0$   & $           1$ & $        1$  \\    
  3  & $              0$   & $          -1$ & $       -1$  \\    
  4  & $              0$   & $          10$ & $       10$  \\    
  5  & $             24$   & $         -14$ & $      -14$  \\    
  6  & $            -24$   & $          85$ & $       93$  \\    
  7  & $            528$   & $        -169$ & $     -201$  \\    
  8  & $           -960$   & $         884$ & $      972$  \\    
  9  & $           8496$   & $       -2390$ & $    -2510$  \\    
 10  & $         -21312$   & $       10212$ & $    10618$  \\    
 11  & $         125904$   & $      -30594$ & $   -31250$  \\    
 12  & $        -380016$   & $      116134$ & $   118792$  \\    
 13  & $        1813416$   & $     -368934$ & $  -378902$  \\    
 14  & $       -6046440$   & $     1337519$ & $  1377207$  \\    
 15  & $       25675200$   & $    -4435616$ & $ -4547052$  \\    
 16  & $      -90096000$   & $    15764526$ & $ 16140138$  \\    
 17  & $      358481304$   & $   -53464296$ & $-54602714$  \\    
 18  & $    -1289158128$   & $   187665313$ &              \\    
 19  & $     4943015520$   & $  -643021360$ &              \\    
 20  & $   -17962232976$   & $  2242649168$ &              \\    
 21  & $    67393016880$   & $ -7729951680$ &              \\    
 22  & $  -245697661872$   & $ 26894409824$ &              \\    
 23  & $   909676085232$   & $-93043627527$ &              \\    
 24  & $ -3315864327216$   &                &              \\
 25  & $ 12172005334848$   &                &              \\
 26  & $-44293518847536$   &                &              \\  \hline
\end{tabular}
\end{center}
\end{table} 
\clearpage
\noindent
and $\{ \lambda_n \}$ by 
\begin{equation}
 \Lambda_2 =  \sum_n \lambda_n u^n.      
\end{equation}
 The latter has been derived from the expansion series 
for $\mu_2$ obtained here 
and the low-temperature series of the susceptibility obtained 
by Guttmann and Enting 
          \cite{Guttmann} 
to order $u^{24}$ 
( They obtained the series to order $u^{26}$ 
and we have checked using the finite-lattice method 
that their coefficients are correct to order $u^{24}$ ). 
 Among the series coefficients obtained here, 
those for $\mu_2$ to order $u^{15}$ 
and those for $\Lambda_2$ to order $u^{12}$ 
coincide to those obtained by Tarko and Fisher 
         \cite{Tarko} 
and we have added new 11 terms.

 We also list the low-temperature series for true correlation length 
squared $\Lambda_2^{\prime}$ 
defined by
         \cite{Tarko} 
\begin{equation}
   \Lambda_2^{\prime}=\frac{1}{ 2 [ \cosh{(m)} -1 ] },
\end{equation}
where  $m$ is the true inverse correlation-length ;
\begin{equation}
     m=
        - \lim_{L \rightarrow \infty}  \frac{1}{L}   
            \log{ \langle {\cal O}(L) {\cal O}(0) \rangle }, 
\end{equation}
where 
$ {\cal O}(\ell)= \sum_{ \{ i | z_i=\ell \} } s_i $ 
is the summation of all the spin variables in $z=\ell$ plane. 
  The coefficients $\{ \lambda_n^{\prime} \}$ are defined by 
\begin{equation}
   \Lambda_2^{\prime} =  \sum_n \lambda_n^{\prime} u^n , 
\end{equation}  
and are derived from the true inverse correlation-length $m$ 
given in Ref.~\cite{Arisuethree_five}.
 We note that, although the coefficients $\{ \lambda_n \}$ differ 
from $\{ \lambda_n^{\prime} \}$  for $n \ge 6$, 
their ratios $\{ \lambda_n^{\prime} / \lambda_n \}$ 
are within a range of 1.02 to 1.03 for  $ 11 \le n \le 17 $. 

\section{Analysis of the series}
\hspace*{\parindent}

    In our preliminary analysis, we estimate the critical exponents
of \( \mu_2 \) and  \( \Lambda_2 \)
by inhomogeneous differential approximants 
               \cite{Hunter_Fisherthree},
in which the approximants to a function \( f(u) \) satisfy 
\begin{equation}
    Q_M(u) f'(u) + P_L(u) f(u) + R_N(u)
    = O(u^{L+M+N+2})                   \label{eqn:Q}.
\end{equation}
 The approximants are equivalent to Dlog Pad\'e approximants 
when \( R_N(u) = 0 \).

\epsfhako{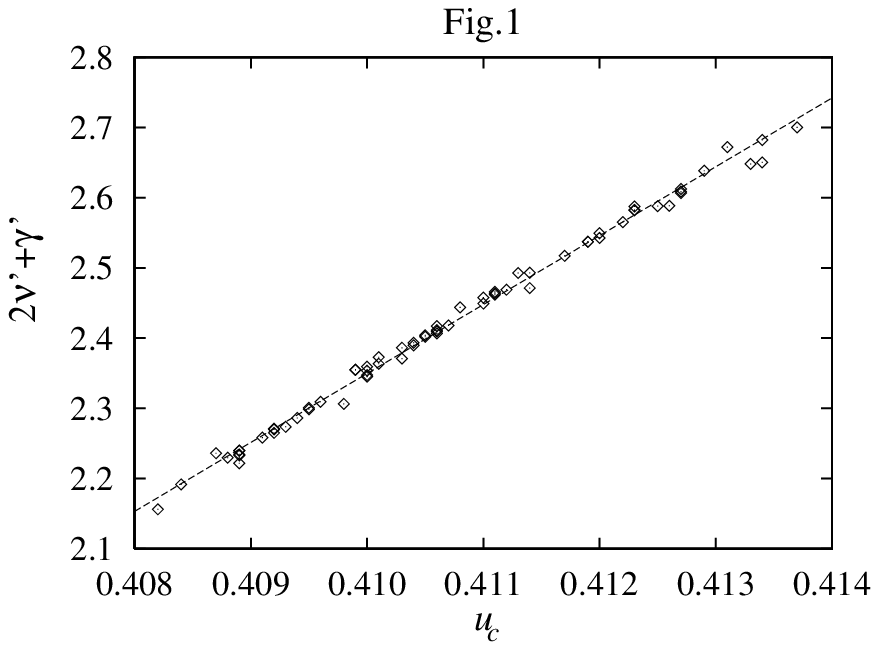}{htb}{10cm}{
  Unbiased estimates of \( 2\nu^{\prime} + \gamma^{\prime} \) 
versus \( u_c \) for \( \mu_2 \);
the data includes the estimates from all the approximants
with \( N = \phi, \ 0, \ 1, \ \cdots , \ 6 \)
and \( L+M+N = 19 \) 
}{fig:fig1}{\vspace{-.5cm}}

 We first give the results of the analysis for \( \mu_2 \). 
 We plot in fig.~\ref{fig:fig1} its critical exponent 
$ 2\nu^{\prime} + \gamma^{\prime} $ 
against the critical point \( u_c \) 
consisting of the data given by all the approximants with
\( L+M+N =19 \) and  \( N = \phi, \ 0, \ 1, \ \cdots, \ 6 \).
 We find a liner correlation 
between the estimates for the critical point and the exponent.
 Linear fitting of these data gives
\( 2\nu^{\prime} + \gamma^{\prime} = 98.1648u_c - 37.8980 \).
 Recent Monte Carlo renormalization-group analysis 
of \( d = 3 \) Ising model 
gives a precise estimate of critical point 
\( u_c = 0.412051 \pm 0.000006 \) 
               \cite{Baillie},
and the series analysis of the high-temperature expansions 
for susceptibility, spontaneous magnetization and specific heat
gives \( u_c \) which is consistent with this
       \cite{Adler_Guttmannthree}.
 Using this value of the critical point, we can read 
from fig.~\ref{fig:fig1} 
that $2\nu^{\prime} + \gamma^{\prime} = 2.5509\pm 0.010 $,
where the error comes from the statistical error in the linear fitting.
 The data given by the approximants with \( N > 6 \)
also fit to almost the same line, but their deviation from the line is 
larger. 

 We plot in fig.~\ref{fig:fig2} 
the estimate of $2\nu^{\prime} + \gamma^{\prime} $ 
obtained by fitting the data from all the approximants
for each fixed \( L+M+N \) 
with \( N = \phi, \ 0, \ 1, \ \cdots, \ 6 \). 
 We note that the estimate appears stable 
for $L+M+N \ge 16$. 
Fitting the data from all the approximants 
with \( 16 \le L+M+N \le 19 \) and 
\( N = \phi, \ 0, \ 1, \ \cdots, \ 6 \)
gives 
\begin{equation}
    2\nu^{\prime} + \gamma^{\prime} = 2.545 \pm 0.012 . 
    \label{eqn:nugamma1}
\end{equation}

\newpage
\epsfhako{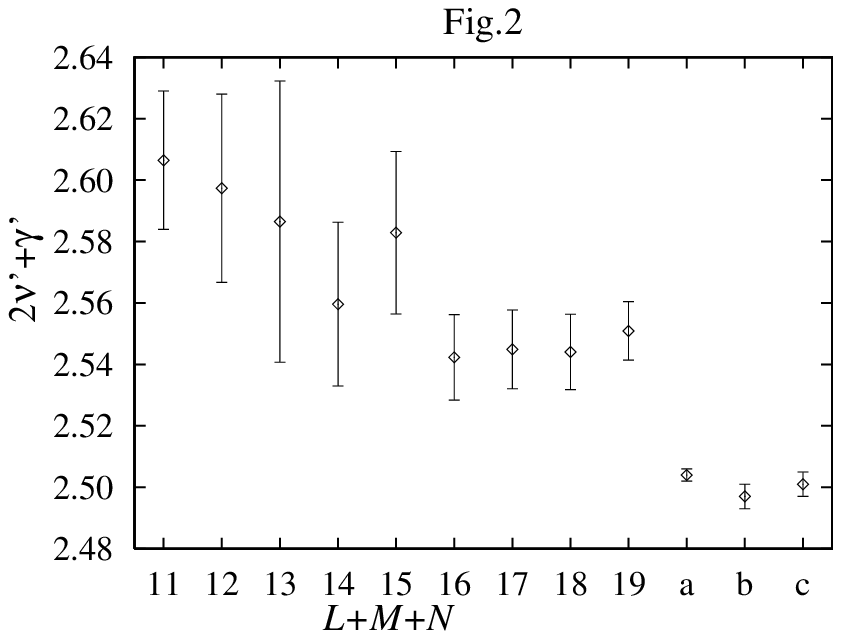}{htb}{10cm}{
  Estimate of \( 2\nu^{\prime} + \gamma^{\prime} \) 
by fitting the data from all the unbiased approximants
with \( N = \phi, \ 0, \ 1, \ \cdots , \ 6 \)
for each fixed \( L+M+N \) ;
 the estimates denoted by (a) and (b) are 
from the high-temperature series 
and (c) from the $\epsilon$-expansion, respectively. 
}{fig:fig2}{\vspace{-.5cm}}
\epsfhako{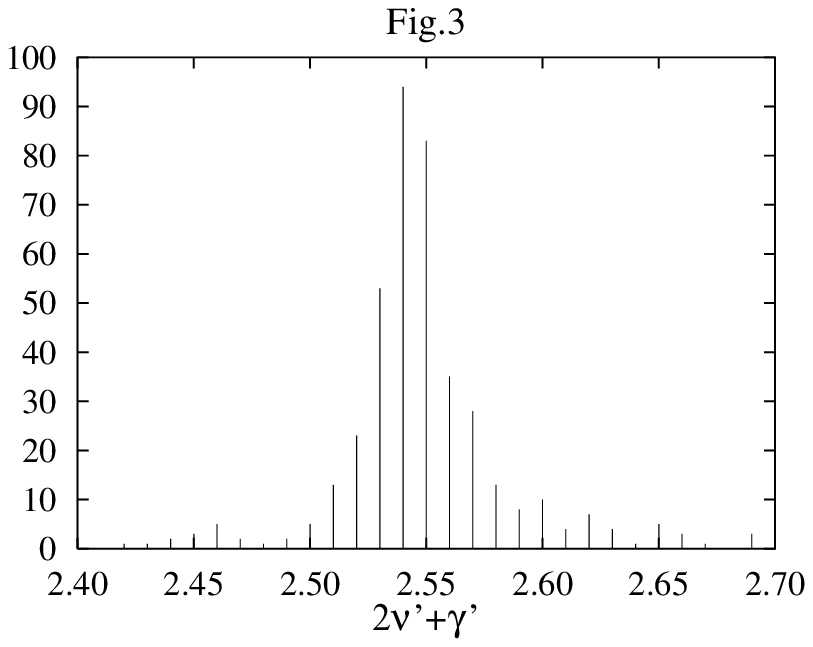}{hb}{10cm}{
   Histogram of the number of biased estimates 
of \( 2\nu^{\prime} + \gamma^{\prime} \)
for \( \mu_2 \);
estimates are obtained from the approximants 
with \( N = \phi, \ 0, \ 1, \ \cdots , \ 6 \) 
and \( 16 \le L+M+N \le 19 \). 
}{fig:fig3}{\vspace{-.5cm}}
\clearpage
\noindent
 We also obtain biased estimates by replacing \( Q_M(u) \)
with \( (u-u_c)Q_M(u) \) in eq.~(\ref{eqn:Q}).
 The estimates of \( 2\nu^{\prime} + \gamma^{\prime} \) show 
a good accumulation around  2.55.
 We show in fig.~\ref{fig:fig3} a histogram of the number of estimates 
given by the biased approximants 
with \( 16 \le L++N \le 19 \) 
and  \( N = \phi, \ 0, \ 1, \ \cdots, \ 6 \). 
 The average and one standard deviation of the data
that satisfy \( |2\nu^{\prime} + \gamma^{\prime} - 2.55| < 0.05 \)
are
\begin{equation}
    2\nu^{\prime} + \gamma^{\prime} = 2.550 \pm 0.018. 
    \label{eqn:nugamma2}
\end{equation}
 
\epsfhako{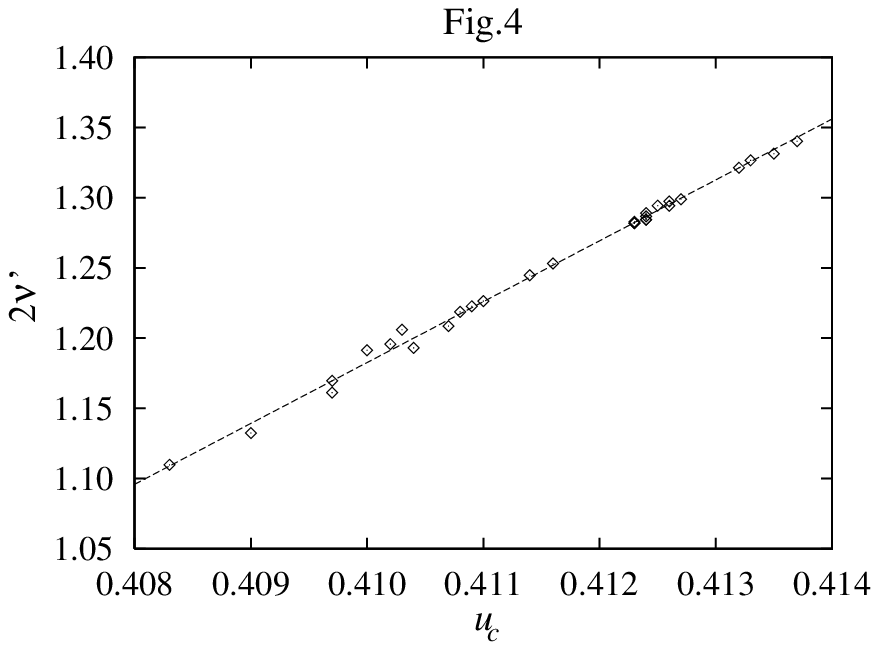}{htb}{10cm}{
    Unbiased estimates of \( 2\nu^{\prime} \) 
versus \( u_c \) for \( \Lambda_2 \);
the data includes the estimates from  all the approximants
with \( N = \phi \)
and \(  17 \le M+L \le 20 \).
}{fig:fig4}{\vspace{-.5cm}}
\begin{table}[htb]
\caption{
      Biased estimates of \( 2\nu^{\prime} \) from \( \Lambda_2 \);
      estimates are obtained by Dlog Pad\'e approximants with
      \( K \equiv M+L = 17,18,19,20 \).
         }
\label{tab:nuprime}
\begin{center}
\begin{tabular}{|c|c|c|c|c|}
\hline
    $[L/M]$   & $K=17$ & $K=18$ & $K=19$ & $K=20$ \\
\hline
$[K-1/1]$   & $0.4495$ &$-0.4837$ & $0.2279$ & $0.1687$ \\
$[K-2/2]$   &$-0.4030$ & $0.8627$ & $2.1319$ & $0.8284$ \\
$[K-3/3]$   & $1.2749$ & $1.2699$ & $1.2794$ & $1.2625$ \\
$[K-4/4]$   & $1.2751$ & $1.2737$ & $1.2731$ & $1.2579$ \\
$[K-5/5]$   & $1.2709$ & $1.2702$ & $1.2731$ & $1.2736$ \\
$[K-6/6]$   & $1.2707$ & $1.2665$ & $1.2708$ & $1.2480$ \\
$[K-7/7]$   & $1.4254$ & $1.2698$ & $1.2707$ & $1.2716$ \\
$[K-8/8]$   & $1.2782$ & $1.2766$ & $1.2815$ & $1.2861$ \\
$[K-9/9]$   & $1.2740$ & $1.2761$ & $1.2778$ & $1.2914$ \\
$[K-10/10]$ & $1.2655$ & $1.2769$ & $1.2735$ & $1.2807$ \\
$[K-11/11]$ & $1.2727$ & $1.2714$ & $1.2488$ & $1.2968$ \\
$[K-12/12]$ & $1.2738$ & $1.2704$ & $1.2725$ & $1.2827$ \\
$[K-13/13]$ & $1.2688$ & $1.2722$ & $1.2735$ & $1.2745$ \\
$[K-14/14]$ & $1.2692$ & $1.2704$ & $1.2637$ & $1.2753$ \\
$[K-15/15]$ & $1.4272$ & $1.2737$ & $1.2675$ & $1.2722$ \\
$[K-16/16]$ & $1.3291$ & $1.2950$ & $1.2889$ & $1.3058$ \\
$[K-17/17]$ &          & $1.2247$ & $1.2879$ & $1.2923$ \\
$[K-18/18]$ &          &          & $1.3850$ & $1.3023$ \\
$[K-19/19]$ &          &          &          & $1.2142$ \\
\hline
\end{tabular}
\end{center}
\end{table} 
 Next we give the result of the analysis for \( \Lambda_2 \). 
The estimates of its critical exponent \( 2\nu^{\prime} \) 
by inhomogeneous differential approximants 
show less convergent results, 
except for the case of \( N = \phi \),
that is, Dlog Pad\'e approximants. 
 In fig.~\ref{fig:fig4} we show the plot of 
the exponent \( 2\nu^{\prime} \) 
against the critical point \( u_c \) 
obtained from Dlog Pad\'e approximants of \( \Lambda_2 \)
with \( 17 \le L+M \le 20 \). 
 Linear fitting of these data gives 
\( 2\nu^{\prime} = 43.360 u_c - 16.595 \).
 We can read from fig.~\ref{fig:fig4} 
using  \( u_c = 0.412051 \) that 
\begin{equation}
    2\nu^{\prime} = 1.272 \pm 0.004.  \label{eqn:nu1}
\end{equation} 
 We also list in table~\ref{tab:nuprime} biased estimates 
of the critical exponent \( 2\nu^{\prime} \) 
by Dlog Pad\'e approximants, 
 and plot in fig.~\ref{fig:fig5} the estimate 
of \( 2\nu^{\prime} \) 
by Dlog Pad\'e approximants for each \( L+M \), 
where we have excluded the data that do not satisfy 
\( | 2\nu^{\prime} - 1.27 | < 0.05 \). 
 All the Dlog Pad\'e approximants with \( 17 \le L+M \le 20 \) 
and  \( | 2\nu^{\prime} - 1.27 | < 0.05 \) 
give
\begin{equation}
    2\nu^{\prime} = 1.276 \pm 0.012. \label{eqn:nu2}
\end{equation}
\epsfhako{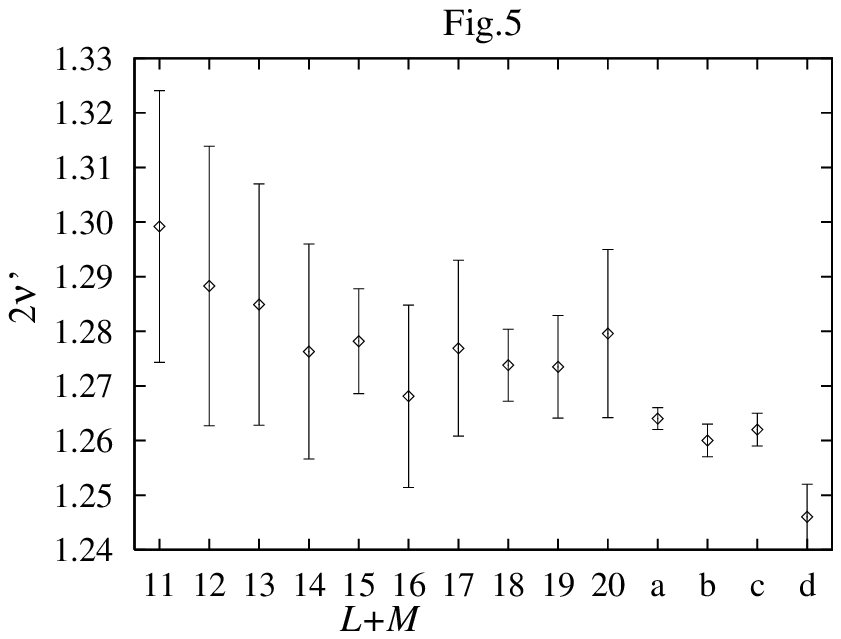}{htb}{10cm}{
  Estimate of \( 2\nu^{\prime} \) 
by biased Dlog Pad\'e approximants
for each fixed \( L+M \);
 the estimates denoted by (a) and (b) are 
from the high-temperature series, 
(c) from the $\epsilon$-expansion, 
and (d) from the Monte Carlo renormalization-group analysis,
respectively. 
}{fig:fig5}{\vspace{-.5cm}}

 Finally we give the result of the analysis for \( \Lambda_2 ' \). 
 The approximants for \( \Lambda_2 ' \) with \( N = 0, \ 1, \ 2 \)
give better converging estimates of \( 2\nu^{\prime} \) than
the other approximants with $N \ge 3$, 
but the average of estimates given by these approximants is too small 
( \( 2\nu^{\prime} \approx 1.20 \) ) 
and inconsistent with the result of the analysis for \( \Lambda_2 \).
( If we use the approximants with \( N \ge 3 \),
we obtain smaller estimates of \( 2\nu^{\prime} \approx 1.17 \). )
 We cannot say that this is due to  
the fact that the series for \( \Lambda_2 ' \)
is shorter than those for \( \Lambda_2 \).
 The analysis of the series for \( \Lambda_2 \) 
to order $u^{17}$, 
which is the maximum order for \( \Lambda_2^{\prime} \), 
gives more convergent and consistent estimates
of \( 2\nu^{\prime} = 1.28 \pm 0.02 \).

 These values of $ 2 \nu^{\prime}+\gamma^{\prime}$ and $2\nu^{\prime} $
are to be compared with the result 
$ 2\nu+\gamma=$ $2.504(2)$, $2.497(4)$ and 
$ 2\nu=       $ $ 1.264(2)$, $1.260(3)$
of the high-temperature series 
          \cite{Fisher,Nickel}, 
$ 2\nu+\gamma=$ $2.501(4)$ and 
$ 2\nu=       $ $ 1.262 (3)$
of the $\epsilon$-expansion
          \cite{Guillou},
and 
$ 2\nu=$      $ 1.246 (6)$
of the Monte Carlo renormalization-group analysis
          \cite{Baillie}.
 The values of $ 2\nu+\gamma$ cited here are estimated 
from the values of $\nu$ and $\gamma$ 
given in the respective references. 
 Our results of $ 2\nu^{\prime}$ 
from \( \Lambda_2 \) [ eqs.~(\ref{eqn:nu1}) and (\ref{eqn:nu2})~]
are not inconsistent with these estimates 
considering the error bounds. 
 On the other hand, 
our results of $ 2\nu^{\prime}+\gamma^{\prime}$ 
from \( \mu_2 \) [eqs.~(\ref{eqn:nugamma1}) 
and (\ref{eqn:nugamma2})~]
are 2 per cent larger 
than the estimates from the high-temperature series 
and $\epsilon$-expansion 
and they are not within error limits. 
 We cannot, however, conclude that there is  
a violation of scaling relation 
$2\nu+\gamma=2\nu^{\prime}+\gamma^{\prime}$, 
considering the fact that our analysis here does not 
include the possiblity of confluent singularity. 
 To take into account the confluent singularity, 
we tried an analysis of Roskies-transformed series 
         \cite{Roskies} 
with the confluent exponent $0.5$, 
but the result was less convergent. 
 It might suggest us to investigate 
inhomogeneous second-order differential approximants
         \cite{Rehr}, 
which is another method to include the confluent singularity.
 Longer series might also solve the discrepancy, 
although the estimate from \( \mu_2 \) appears so stable 
when we change $L+M+N$ from $16$ to $19$ 
in our analysis of inhomogeneous differential approximants 
as mensioned above ( See fig.~\ref{fig:fig2}). 

 We find the unphysical but dominant singularity
at \( u = u_1 = -0.2858 \pm 0.0003 \) 
with a critical exponent
\begin{equation}
    2\nu^{\prime} + \gamma^{\prime}\rm{(unphysical)} = 1.892 \pm 0.034, 
\end{equation}
from \( \mu_2 \)
and
at \( u = u_1 = -0.2858 \pm 0.0006 \) 
with a critical exponent
\begin{equation}
    2\nu^{\prime} \rm{(unphysical)} = 0.811 \pm 0.039, 
\end{equation}
from \( \Lambda_2 \).
 These values of the critical point are consistent 
with \( u_1 = -0.2853(3) \) 
from the susceptibility 
          \cite{Guttmann}, 
and the critical exponents are 
so sensitive to the value \( u_1 \) of the critical point 
if we would try a biased estimation, 
in which the change of the position of the critical point by $0.0005$ 
reduces the critical exponent 
about 0.08 for $\mu_2$ and about 0.03 for $\Lambda_2$.

\section{Summary}
\hspace*{\parindent}
 
  We have calculated low-temperature series 
for the second moment of the correlation function 
in $d=3$ simple-cubic Ising model 
to order $u^{26}$ by finite-lattice method, 
from which we have obtained the low-temperature series 
for the second-moment correlation length to order $u^{23}$ 
using the known low-temperature series for the susceptibility. 
 The obtained series is 11 terms longer 
than those calculated previously.
  Preliminary analysis of the series 
by inhomogeneous differential approximants 
gives critical exponents  
$ 2\nu^{\prime} + \gamma^{\prime} \approx 2.55 $ 
and 
$ 2\nu^{\prime} \approx 1.27 $. 
 The latter is not inconsistent with 
the result from high-temperature series and $\epsilon$-expansion, 
but there is a discrepancy by 2 per cent 
between the former and the critical exponent 
from high-temperature series and $\epsilon$-expansion.
 It appears that further analysis of the series 
should be done including the possibility of the confluent singularity. 

\section*{Acknowledgements}
\hspace*{\parindent}

 We would like to thank M. E. Fisher for calling our attention  
to the results of the critical exponents in recent articles 
and G. M\"unster for valuable discussions on the finite-lattice method.


\newpage
\begin{thebibliography}{99}
\bibitem{Arisuethree_five}
    H. Arisue and T. Fujiwara, 
      Nucl. Phys. B285[FS19] (1987) 253; \\
    H. Arisue and K. Tabata, 
      Phys. Letters B322 (1994) 224
\bibitem{Bhanot}
    G. Bhanot, M. Creutz, J. Lacki, 
      Phys. Rev. Lett. 69 (1992) 1841
\bibitem{Guttmann}
    A. J. Guttmann and I. G. Enting, 
      J. Phys. A26 (1993) 807
\bibitem{Arisuefour}
    H. Arisue, 
      Phys. Letters B313 (1993) 187
\bibitem{Neef}
    T. de Neef and I. G. Enting, 
      J. Phys. A10 (1977) 801
\bibitem{Arisueone_two_six}
    H. Arisue and T. Fujiwara, 
      Prog. Theor. Phys. 72 (1984) 1176; \\
    H. Arisue and T. Fujiwara, 
      Preprint RIFP-588 (1985 unpublished); \\
    H. Arisue, 
      Nucl. Phys. B (Proc. Suppl.) 34 (1994) 240
\bibitem{Creutz}
    M. Creutz, 
         Phys. Rev. B43 (1991) 10659
\bibitem{Tarko}
    H. B. Tarko and M. E. Fisher, 
      Phys. Rev. B11 (1975) 1217
\bibitem{Muenster}
  G. M\"unster, 
         Nucl. Phys. B180[FS2] (1981) 23
\bibitem{Bhanottwo}
  G. Bhanot,
         J. Stat. Phys. 60 (1990) 55
\bibitem{Hunter_Fisherthree}
    G. L. Hunter and G. A. Baker,
      Phys. Rev. B19 (1979) 3808; \\
    M. E. Fisher and H. Au-Yang,
      J. Phys. A12 (1979) 1677
\bibitem{Baillie}
    C. F. Baillie, R. Gupta, K. A. Hawick and G.S. Pawley,
         Phys. Rev. B45 (1992) 10438
\bibitem{Adler_Guttmannthree}
    J. Adler, 
      Phys. Rev. B36 (1987) 2473;  \\
    A. J. Guttmann, 
      J. Phys. A20(1987) 1855
\bibitem{Fisher}
    M. E. Fisher and J. H. Chen,
      J. Physique  46 (1985) 1645
\bibitem{Nickel}
    B. G. Nickel and J. J. Rher, 
      J. Stat. Phys. 61 (1990) 1
\bibitem{Guillou}
    J. C. Le Guillou and J. Zinn-Justin, 
      J. Physique 48 (1987) 19
\bibitem{Roskies}
  R. Z. Roskies,
      Phys. Rev. B24 (1981) 5305
\bibitem{Rehr}
  J. J. Rehr, G. S. Joyce and A. J. Guttmann, 
      J. Phys. A13 (1980) 1587
\end{thebibliography}
\end{document}